# Microscopic quantum ideal rotor model and related self-consistent cranking model, I: uni-axial rotation case

P. Gulshani

NUTECH Services, 3313 Fenwick Crescent, Mississauga, Ontario, Canada L5L 5N1
Tel. #: 647-975-8233; matlap@bell.net

A microscopic quantum ideal rotor-model Hamiltonian (distinct from that of Bohr's rotational model) is derived for a rotation about a single axis by applying a dynamic rotation operator to the deformed nuclear ground-state wavefunction. It is shown that the microscopic ideal rotor Hamiltonian is obtained only for a rigid-flow prescription for the rotation angle, with the attendant rigid-flow kinematic moment of inertia. (For the case of the center-of-mass motion, the method predicts the correct mass.) Using Hartree-Fock variational and second quantization methods, the ideal rotor-model Hamiltonian is reduced to that of a self-consistent cranking model plus residual terms associated with the square of the angular momentum operator and a two-body interaction. The approximations and assumptions underlying the conventional cranking model are revealed. The resulting nuclear Schrodinger equation, including a residual two-body interaction and the residual part of the square of the angular momentum, is then solved in the Tamm-Dancoff approximation using the eigenstates of the self-consistent cranking model, with a self-consistent deformed harmonic oscillator potential, as the particle-hole basis states. Good agreement is obtained between the predicted and measured ground-state rotational-band excitation energies, including the lowering of the excitation energy with increasing angular momentum, in $^{20}_{10}Ne$ when the effects of a 3-D rotation are simulated in the model.

*PACS number: 21.60.Ev, 21.60.Fw, 21.60.Jz*

## 1. Introduction

The phenomenological self-consistent conventional semi-classical cranking model (*CRM*) with a constant angular velocity [1,2] is frequently and successfully used [3-32 and references therein] to predict rotational properties and phenomena in deformed nuclei. It is therefore desirable to derive this model from first principles and reveal the assumptions and approximations that underlie it. There have been many studies to achieve this objective using various methods, approximations and assumptions [2-4,7,8,17,24,33,34,36-53]. In this article, we use a simple approach to obtain a microscopic quantum ideal rotor model Hamiltonian, which is then reduced to a self-consistent cranking model Hamiltonian plus residual quantum microscopic corrections. It is shown that the microscopic ideal rotor model Hamiltonian derived here is distinct from that of the Bohr's rotational model.

The conventional cranking model for rotation about a single axis assumes that the anisotropic nuclear potential $V$ is rotating at a constant angular frequency $\omega_{cr}$ about $x$ or 1 axis. The model time-dependent Schrodinger equation is[1]:

$$i\hbar \frac{\partial}{\partial t}|\Psi_{cr}\rangle = \hat{H}|\Psi_{cr}\rangle \quad (1)$$

where:

$$\hat{H} \equiv \frac{1}{2M}\sum_{n,j=1}^{A,3} \hat{p}_{nj}^2 + \hat{V}_{cr}(\vec{r}_n), \qquad \vec{r}_n = R(\omega_{cr}t)\cdot\vec{r}_n', \quad (2)$$

$R$ is an orthogonal matrix and $\vec{r}_n'$ is the $n^{th}$ particle coordinate relative to the rotating frame. Eq. (1) is unitarily transformed to the rotating frame:

$$|\Psi_{cr}\rangle = e^{-i(\omega_{cr}\hat{J}+E)t/\hbar}|\Phi_{cr}\rangle \quad (3)$$

where $\hat{J}$ is the $x$-component of the total angular momentum operator. One then obtains the stationary *CRM* equation[2]:

$$\hat{H}_{cr}\cdot|\Phi_{cr}\rangle \equiv (\hat{H}-\omega_{cr}\cdot\hat{J})\cdot|\Phi_{cr}\rangle = E_{cr}|\Phi_{cr}\rangle \quad (4)$$

The angular velocity $\omega_{cr}$ is determined as a function of $J$ by requiring the expectation of $\hat{J}$ to have the fixed value $\hbar J$:

$$\hbar J \equiv \langle\Phi_{cr}|\hat{J}|\Phi_{cr}\rangle \quad (5)$$

The excited-state rotational energy $E_J$ in a space-fixed frame is then given by:

$$E_J = \langle\Phi_{cr}|\hat{H}|\Phi_{cr}\rangle = \langle\Phi_{cr}|(\hat{H}_{cr}+\omega_{cr}\cdot\hat{J})|\Phi_{cr}\rangle = E_{cr} + \omega_{cr}\cdot\langle\Phi_{cr}|\hat{J}|\Phi_{cr}\rangle \quad (6)$$

The effective dynamical moment of inertia $\mathcal{I}_{eff}$ is not an observable and must be deduced from other predicted or measured nuclear properties. A definition of $\mathcal{I}_{eff}$, which is adopted from rigid-body rotation and is commonly used, is given at each value of $J$ by the excitation energy $\Delta E_J$:

$$\frac{2\mathcal{I}_{eff}}{\hbar^2} = \frac{4J-2}{\Delta E_J - \Delta E_{J-2}} \quad (MeV)^{-1} \quad (7)$$

$$\Delta E_J \equiv E_J - E_0 \quad (8)$$

In Section 2 of this article, we derive from first principles an ideal rotor model (that is distinct from that of Bohr rotational model) for a rotation about a single axis by applying a dynamic rotation operator to a deformed nuclear ground-state (or excited band-head) wavefunction and thereby obtain the corresponding transformed (rotor) nuclear Hamiltonian[3]. In this application, we distinguish between static and dynamic types of rotation and their impact on the results. The rotation angle is chosen to be defined by a rigid-flow component of the nucleon velocity field. This definition eliminates coupling terms between the angular momentum and other operators in the transformed Hamiltonian, giving the rotor Hamiltonian a purely (ideally) intrinsic or non-rotational character and containing only the square of the angular momentum. For the center-of-mass motion, the method predicts the correct mass.

---

[1] Clearly, this time-dependent description of the rotational motion is classical in nature. Furthermore, the *c*-number parameter $\omega_{cr}$ is not an operator acting on a nucleon probability distribution.

[2] Eq. (4) can also be derived from a variation of the Schrodinger equation subject to energy minimization, with the wavefunction $\Phi_{cr}$ constrained to give a fixed value for the expectation of the angular momentum operator.

In Section 3, the rotationally and time-reversal invariant rotor Hamiltonian is reduced to a self-consistent cranking model using Hartree-Fock (HF) mean-field variational second quantization methods, and an expression is derived for the remaining part (i.e., the HF direct and exchange parts of the one-body and two-body parts) of the square of the angular-momentum operator, which is then treated as a residual interaction. The approximations and assumptions underlying the conventional cranking model are identified.

In Section 4, the HF self-consistent cranking model equation is solved in a closed form using a self-consistent deformed harmonic oscillator potential and an isotropic velocity distribution constraint (similar to that used in [13]) to satisfy the rigid-flow constraint mentioned above. A comparison of the predicted and measured excitation energies is presented for $^{20}_{10}Ne$.

In Section 5, to remove a discrepancy between the excitation energy observed in $^{20}_{10}Ne$ and that predicted by the HF self-consistent cranking model, we solve the microscopic ideal rotor-model Schrodinger equation, including the residual angular-momentum operator and a residual schematic two-body interaction, in the Tamm-Dancoff approximation using the cranked HF states derived in Section 4 as particle-hole basis states.

In Section 6, the microscopic ideal rotor model of Section 5 is used to predict the ground-state rotational-band excitation energy and quadrupole moment in $^{20}_{10}Ne$ and the results are compared with the corresponding measured data. To obtain a good fit to the data including lowering of the rotational-state energy with increasing angular momentum, the solution of the rotor-model equations is modified to include the impact of a 3-D rotation. The results are discussed and compared with the results from other related studies.

Section 7 concludes the article.

## 2. Derivation of microscopic quantum ideal rotor model for un-axial rotation

To derive the microscopic quantum ideal rotor model (which is distinct from Bohr's rotational model) for a rotation about a single axis[4], we consider a deformed distribution of nucleons in the nuclear ground state described by the wavefunction $|\Phi_{gs}\rangle$ obtained by some method such as HF. We assume that $|\Phi_{gs}\rangle$ is an approximate ground state of the nucleus and hence satisfies approximately the nuclear Schrodinger equation for a rotationally-invariant Hamiltonian $\hat{H}_o$:

$$\hat{H}_o|\Phi_{gs}\rangle = E_{gs}|\Phi_{gs}\rangle, \qquad \hat{H}_o = \sum_{n=1}^{A}\frac{\hat{p}_n^2}{2M} + \frac{1}{2}\sum_{n,m}^{A}\hat{V}(|\vec{r}_n - \vec{r}_m|) \qquad (9)$$

where $\hat{V}$ (spin, isospin, and exchange dependence of $V$ is left out for now) is a rotationally invariant two-body interaction, and $A$ is the mass number. Next we rotate the deformed ground-state wavefunction $|\Phi_{gs}\rangle$ (i.e., the deformed nucleon distribution) through a fixed angel $\theta$ about the *x*-axis to obtain a rotated state $|\Phi(\theta)\rangle$ as follows:

$$|\Phi(\theta)\rangle = e^{i\theta\hat{J}/\hbar}\cdot|\Phi_{gs}\rangle \qquad (10)$$

---

[3] This method is a refinement and generalization of those used previously in [38,39,41,42,43], and is somewhat similar to the angular momentum projection method [33,36 and references therein] but differs from it in the following ways: we do not use angular momentum projection and expansion in angular momentum, we use simplifying rigid-flow prescription for the rotation angle exploiting the commutator of the angle and angular momentum operators, we derive a purely intrinsic cranked Hamiltonian, and we derive a self-consistent cranking model supplemented by residual angular-momentum operator and two-body interaction.

[4] The restriction of the rotation to one spatial dimension is of classical nature. It is adopted here from the conventional cranking model because one of the objectives here is to drive a quantum mechanical analogue of the cranking model. This classical feature will be removed when the microscopic model is generalized to 3-D rotation.

In Eq. (10), the rotation generated by the rotation operator $e^{i\theta\hat{J}/\hbar}$ is static (as is normally used in the literature for single and multi-particle systems [7,11,17,33,36, and references therein]), i.e., $\theta$ is a constant parameter and hence it commutes with the angular momentum operator $\hat{J}$ along the *x* axis, so that the operation in Eq. (10) is purely rotational in nature. Otherwise (i.e., if we regard $\theta$ and $\hat{J}$ to be a canonically conjugate operators in the rotation operation in Eq. (10)) the operation in Eq. (10) would become a combination of distortion and rotation of the nucleon distribution. And would generate unphysical results to appear in the transformed Schrodinger equation, such as a kinematic moment of inertia larger than that for a rigid flow (although these unphysical results can be eliminated by a simple modification of the rotation operator that allows a desired degree of non-commutativity of $\theta$ and $\hat{J}$).

However, for each given orientation of the nucleon spatial distribution specified by $\theta$, we consider $\theta$ to be a function of the nucleon co-ordinates[5]. Since $\hat{J}$ is the generator of infinitesimal rotation of the nucleon position vectors, we may require $\theta$ to be canonically conjugate to $\hat{J}$, obeying the commutation relation:

$$[\hat{\theta},\hat{J}] = i\hbar \qquad (11)$$

Eq. (11) partially determines the functional dependence of $\theta$ on the nucleon co-ordinates, refer to Eqs. (14) and (15).

Substituting Eq. (10) for $|\Phi_{gs}\rangle$ into the first of Eqs. (9), we obtain the transformed Schrodinger equation:

$$\hat{H}|\Phi\rangle = E_{gs}|\Phi\rangle \qquad (12)$$

where the transformed nuclear Hamiltonian $\hat{H}$ is:

---

[5] $\theta$ is not explicitly a function of the nucleon spin. However, since the nucleon spatial distribution is determined by the intrinsic wavefunction $|\Phi_{gs}\rangle$, which depends on the nucleon spin, $\theta$ depends indirectly on the spin. The rotational model developed here is valid for any momentum-independent nuclear interaction and for a system of fermions or bosons, depending on whether the intrinsic wavefunction is anti-symmetrized or symmetrized respectively.

$$\hat{H} \equiv e^{i\theta \hat{J}/\hbar} \cdot \hat{H}_o \cdot e^{-i\theta \hat{J}/\hbar} = \hat{H}_o + \frac{i}{2\hbar M} \cdot \sum_{n=1}^{A}\left[\theta, \hat{p}_n^2\right] \cdot \hat{J} - \frac{1}{4\hbar^2 M} \sum_{n=1}^{A}\left[\theta,\left[\theta, \hat{p}_n^2\right]\right] \cdot \hat{J}^2 \quad (13)$$

which is readily derived using the commutator expansion:

$$e^A B e^{-A} = B + [A,B] + \frac{1}{2!}[A,[A,B]] + \frac{1}{3!}[A,[A,[A,B]]] + \cdots.$$

for any operators $A$ and $B$. Eq. (12) shows that each rotated wavefunction $|\Phi(\theta)\rangle$ in Eq. (10) satisfies Eq. (12) with the same ground-state energy $E_{gs}$. That is, the Hamiltonian $\hat{H}$ and $|\Phi(\theta)\rangle$ for all orientations $\theta$ describe degenerate (or collapsed) rotational states with the same energy $E_{gs}$. Hence, we may conclude that $\hat{H}$ is an intrinsic Hamiltonian and $|\Phi(\theta)\rangle$ is a superposition of angular momentum eigenstates. (For the case of the center-of-mass motion, Eq. (13) becomes: $\hat{H} = H_o - \frac{1}{2MA}P^2$, where $P$ is the center-of-mass momentum, and hence the correct mass $MA$ is predicted.)

The rotation angle $\theta$ can be chosen arbitrarily, and the final calculated results will be the same. We define $\theta$ in terms of the nuclear quadrupole distribution since experimental and theoretical observations indicate that nuclear rotational motion is dominated by that of the quadrupole nucleon distribution (Bohr-Mottelson's quadrupole deformation model and numerous other collective models such as Villars' collective models using quadrupole moment to define the rotation angle are testaments to this fact). In line with this observation, to obtain the simplest possible and non-trivial expression for $\hat{H}$ in Eq. (13), and to ensure that $\hat{H}$ is a purely (ideally) intrinsic operator in view of the appearance of the second term on the right-hand side of Eq. (13), we define $\theta$ as follows:

$$\frac{\partial \theta}{\partial x_{nj}} = -\sum_{k=1}^{2} \chi_{jk}\, x_{nk}, \quad \chi_{jk} = -\chi_{kj} = 0 \text{ for } j,k \neq 2,3 \quad (14) \text{ where}$$

$\chi$ is a real 3x3 anti-symmetric matrix. The choice in Eq. (14) adds a collective rigid-flow component to each nucleon velocity field, and it renders the second term on the right-hand-side of Eq. (13) quadratic in $\hat{J}$. The non-zero element $\chi_{23}$ of $\chi$ is determined by substituting Eq. (14) into Eq. (11) to obtain:

$$\chi_{23} = \frac{1}{\hat{\mathcal{I}}_+}, \qquad \hat{\mathcal{I}}_+ \equiv \sum_{n=1}^{A}\left(y_n^2 + z_n^2\right) \quad (15)$$

where $M\hat{\mathcal{I}}_+$ is the rigid-flow moment of inertia[6]. (Note that for any other choice of $\theta$ such as rigid-plus-irrotational flow prescription used in [41,42,43], the second term on the right-hand side of Eq. (13) will have, in addition to a $\hat{J}^2$ term, terms with $\hat{J}$ coupled to other types of operators such as a shear operator, complicating $\hat{H}$ properties.)

---

[6] Note that the rigid-flow prescription for $\theta$ in Eqs. (14) and (15) is a collective analogue of the Birbrair's single-particle $\theta_n$ [54]: $\vec{\nabla}_n \theta_n = \vec{e}_x \times \vec{r}_n / (y_n^2 + z_n^2)$, where $\vec{e}_x$ is a unit vector along the x axis. $\theta_n$ has continuous second-order mixed derivatives (i.e., $\vec{\nabla}_n \times \vec{\nabla}_n \theta_n = 0$) in any spatial region that excludes the x axis, along which $\theta_n$ is singular. Whereas $\theta$ has discontinuous second-order mixed derivatives (i.e., $\vec{\nabla}_n \times \vec{\nabla}_n \theta \neq 0$). The difference in the discontinuity between $\vec{\nabla}_n \theta$ and $\vec{\nabla}_n \theta_n$ arises because of the many-body nature of $\hat{\mathcal{I}}_+^{-1}$ in Eqs. (14) and (15).

Inserting Eq. (14) into Eqs. (12) and (13), we obtain:

$$\hat{H}|\Phi\rangle = \left(\hat{H}_o - \frac{\hat{J}^2}{2M\hat{\mathscr{I}}_+}\right)|\Phi\rangle = E_{gs}|\Phi\rangle \tag{16}$$

We require the wavepacket $|\Phi(\theta)\rangle$ to describe a state with a given average angular momentum $\hbar J$, i.e., to satisfy the angular-momentum constraint (as in the conventional cranking model):

$$\hbar J = \langle \Phi|\hat{J}|\Phi\rangle \tag{17}$$

because $|\Phi\rangle$ is not an eigenstate of $\hat{J}$ but rather a superposition of such states. Note that, for uni-axial rotation, we use $\hbar J$ in Eq. (17) instead of $\hbar\sqrt{J(J+1)}$ to obtain the correct cut-off angular momentum value when the system becomes axially symmetric about the rotation axis in the cranking model derived in Section 3.

We mentioned above that $\hat{H}$ in Eq. (16) may be interpreted as an intrinsic Hamiltonian for a system with the kinematic moment of inertia $M\hat{\mathscr{I}}_+$. Indeed, using Eq. (16), we find that $\hat{H}$ in Eq. (16) is a purely (ideally) intrinsic Hamiltonian, because it satisfies the condition[7]:

$$\left[\hat{H},\theta\right] = 0 \tag{18}$$

Indeed, in many studies, the operator $\hat{J}^2/\hat{\mathscr{I}}$ with some inertia parameter $\hat{\mathscr{I}}$ has been used to remove spurious rotational-energy excitations. The above results then show that for this removal to be exactly valid, $\hat{\mathscr{I}}$ must be replaced by $M\hat{\mathscr{I}}_+$. A Hamiltonian of the form of $\hat{H}$ in Eq. (16) with an arbitrary inertia parameter for $\hat{\mathscr{I}}_+$ has been used in many studies of nuclear collective rotation [44-53,55-59].

Eq. (16) can be expressed as follows:

$$\hat{H}_o|\Phi\rangle = \left(\hat{H} + \frac{\hat{J}^2}{2M\hat{\mathscr{I}}_+}\right)|\Phi\rangle \equiv E_J|\Phi\rangle \tag{19}$$

Eq. (19) resembles that in the Bohr's rotational model [7,11,12,14,17,38,39,60] but differs inherently from it in the following ways: $|\Phi\rangle$ is not a product of an angular momentum eigenstate and an intrinsic state but rather is a superposition of angular momentum eigenstates (i.e., it is not a product of rotation and intrinsic wavefunctions as in the Bohr's rotational model), and $M\hat{\mathscr{I}}_+$ is a kinematic rather than a dynamic moment of inertia even though $\hat{H}$ an intrinsic Hamiltonian as in the Bohr's model. Therefore, Eq. (19) and Bohr's corresponding equation are

---

However, the discontinuity in $\vec{\nabla}_n\theta$ is small because the expectation of $\hat{\mathscr{I}}_+$ is a large number. On the other hand, for a rigid-body-type of motion $\vec{\nabla}_n\theta_{rig} = \omega_{rig}\vec{e}_x \times \vec{r}_n$, where angular velocity $\omega_{rig}$ is a constant, $\vec{\nabla}_n \times \vec{\nabla}_n\theta_{rig} \neq 0$, i.e., $\vec{\nabla}_n\theta_{rig}$ is inherently discontinuous. We note that second-order mixed derivatives of $\theta$ do not appear anywhere in the derivation of the equations in this article.

[7] The conditions in Eqs. (11) and (18) are similar to those for a Goldstone boson or phonon arising in an RPA mode with zero excitation energy and zero restoring force, and identified with a rotational motion [47,49-53].

not expected to predict similar results. In fact, from an analysis of angular momentum projection of deformed HF states, Bouten-Caurier [60] have concluded that the rotational motion predicted by Bohr's rotational model is not applicable to the rotational motion in the light nuclei, except at very large deformation, whereas the conventional cranking model has been largely successful in predicting rotational spectra in the light nuclei. The above results and the success of the cranking model in predicting nuclear rotational properties imply that a clean and complete separation of rotation and intrinsic motions in nuclei may not be possible (except at very large deformation). This conclusion may also be supported by the analyses in [41,42,43] where it was found that such a separation was not possible except under an extreme condition.

It may be of interest to note that Eq. (19) may be related to the procedure used by Skyrme-Levinson-Kelson, et. al., [55-58] in computing an approximate average value of the dynamic moment of inertia.

### 3. Deriving HF cranking model plus residual operators from Eq. (16)

In this section, we reduce Eq. (16) to that of a self-consistent version of the conventional cranking model Eq. (4) plus the remaining correction terms. We do this using HF variational and second quantization methods.

The inverse of the kinematic rigid-flow moment of inertia $M\hat{\mathcal{I}}_+$ in Eq. (16) is a many-body operator. Therefore, to make the calculation tractable, and since we are concerned only with rotational, and not vibrational, motion, we replace $\hat{\mathcal{I}}_+$ by its expectation $\mathcal{I}_+^o$ over the state $|\Phi\rangle$ (noting that $\mathcal{I}_+^o$ is large and varies very gradually with $J$). Eq. (16) then becomes:

$$\hat{H}|\Phi\rangle = \left(\hat{H}_o - \frac{\hat{J}^2}{2M\mathcal{I}_+^o}\right)|\Phi\rangle = E_{gs}|\Phi\rangle \qquad (20)$$

where:

$$\mathcal{I}_+^o \equiv \langle\Phi|\hat{\mathcal{I}}_+|\Phi\rangle \qquad (21)$$

Applying the HF variational principle to Eq. (20)[8], we obtain the following single-particle self-consistent cranking model equation:

$$\hat{h}|\varphi_n\rangle \equiv \left(\hat{h}_o - \omega\cdot\hat{j}\right)|\varphi_n\rangle = \varepsilon_n|\varphi_n\rangle \qquad (22)$$

where $\hat{h}_o$ and $\omega\cdot\hat{j}$ are the single-particle direct HF mean-field part of nuclear Hamiltonian $\hat{H}_o$ and $\hat{J}^2/2M\mathcal{I}_+^o$ respectively, and the rotation angular velocity $\omega$ is defined by:

$$\omega M\mathcal{I}_+^o \equiv \langle\Phi|\hat{J}|\Phi\rangle = \hbar J \qquad (23)$$

where we have used the angular-momentum constraint in |Eq. (17).

---

[8] In this variation, the contribution from the change in $\mathcal{I}_+^o$ is neglected because it is small compared to that in the expectation of $\hat{J}^2$ since $\mathcal{I}_+^o$ is large and varies little with $J$ or $\omega$. For these reasons, we replace $\mathcal{I}_+^o$ in Eq. (23) by $\mathcal{I}_+^o = \langle\Phi_{crHF}|\hat{\mathcal{I}}_+|\Phi_{crHF}\rangle$ where $|\Phi_{crHF}\rangle \equiv \hat{P}\prod_{n=1}^{A}\varphi_n$ and $\hat{P}$ is anti-symmetrization operator.

The rigid-flow condition in Eq. (23) ensures that the Hamiltonian $\hat{H}$ in Eq. (20) is purely an intrinsic quantity as defined in Eq. (18). This condition is in addition to that required by the HF mean-field approximation (such as minimization of the mean-field energy subject to a constant volume when an approximate potential is used as a substitute for the actual HF mean-field potential).

The microscopic self-consistent cranking model Eqs. (22) and (23) become identical to the conventional cranking model Eqs. (4) and (5) when the moment of inertia $M\mathcal{S}_+^o$ in Eq. (23) is replaced by an arbitrary inertia parameter. But then the Hamiltonian $\hat{H}$ in Eq. (20) would no longer be a purely intrinsic quantity. Therefore, the conventional cranking-model Hamiltonian in Eq. (4) is not a purely intrinsic Hamiltonian because it ignores the rigid-flow condition in Eq. (14), in addition to ignoring the HF exchange term of the one-body part and other residual parts of the square of the angular momentum operator.

To regain the Hamiltonian $\hat{H}$ in Eq. (20), we must add to the HF mean-field Eq. (22), the remaining or residual parts of the nuclear interaction $V$ in $\hat{H}_o$ (refer Eq. (9)) and $\hat{J}^2$ in Eq. (20). The residual part of $\hat{J}^2$ is given in the second-quantized representation by:

$$\left(\hat{J}^2\right)_{res} \equiv \hat{J}^2 - 2\langle\hat{J}\rangle\cdot\hat{J}$$
$$= \sum_{\nu\nu'}\left[\left(\hat{j}^2\right)_{\nu\nu'} - 2\sum_{k=1}^{A}\hat{j}_{\nu k}\cdot\hat{j}_{k\nu'}\right]\cdot a_\nu^\dagger a_{\nu'} + \sum_{l,k=1}^{A}\hat{j}_{lk}\cdot\hat{j}_{kl} - \langle\hat{J}\rangle^2 + \sum_{\mu\mu'\nu\nu'} j_{\mu\mu'}\cdot j_{\nu\nu'} : a_\mu^\dagger a_\nu^\dagger a_{\nu'} a_{\mu'}: \quad (24)$$

where the subscripts indicate matrix elements between the (HF) self consistent cranked orbitals in Eq. (22), $\mu$ and $\nu$ range over occupied and unoccupied cranked HF orbitals, angled brackets indicate expectation over cranked HF nuclear ground state, and colons indicate normal ordering with respect to the cranked HF ground state. The first two terms in Eq. (24) are the direct and exchange HF parts of the one-body part of $\hat{J}^2$. The next term in Eq. (24) is the HF expectation of one-body part of $\hat{J}^2$. The last term in Eq. (24) is the residual of the two-body part of $\hat{J}^2$.

We now add to the HF mean-field Schrodinger Eq. (22) the residual part $\hat{V}_{res}$ of the two-body interaction $V$ in $\hat{H}_o$ in Eq. (9), and the residual part of the rotational kinetic energy in Eq. (20). Eq. (20) then becomes:

$$\hat{H}|\Phi\rangle = \left[\hat{H}_{crHF} + \hat{V}_{res} - \frac{\left(\hat{J}^2\right)_{res}}{2M\mathcal{S}_+^o}\right]|\Phi\rangle = E_{gs}|\Phi\rangle \quad (25)$$

where

$$\hat{H}_{crHF}|\Phi_{crHF}\rangle \equiv \sum_{n=1}^{A}\left[\hat{h}_o(n) - \omega\cdot\hat{j}(n)\right]|\Phi_{crHF}\rangle = E_{gsHF}|\Phi_{crHF}\rangle \quad (26)$$

$$|\Phi_{crHF}\rangle \equiv \hat{P}\prod_{n=1}^{A}|\varphi_n\rangle, \quad E_{gsHF} \equiv \sum_{n=1}^{A}\varepsilon_n \quad (27)$$

$|\varphi_n\rangle$ and $\varepsilon_n$ are given by Eq. (22), and $\left(\hat{J}^2\right)_{res}$ is given in Eq. (24). For $\hat{V}_{res}$ we choose the separable effective quadrupole-quadrupole residual interaction (which is often used in nuclear structure calculations):

$$\hat{V}_{res} = -\frac{\chi}{2} Q^\dagger \cdot Q, \qquad \hat{Q}_\mu \equiv \sum_{k=1}^{A} r_k^2 \hat{q}_\mu \equiv \sum_{k=1}^{A} r_k^2 \cdot Y_{2\mu}(\theta, \phi), \qquad \mu = 0, \pm 1, \pm 2 \qquad (28)$$

where the parameter $\chi$ is the interaction strength.

## 4. Solving self-consistent cranking model Eqs. (22) and (23)

In this section, we solve the self-consistent cranking model (*SCRM*) Schrodinger Eqs. (22) and (23) for a self-consistent (mean-field) deformed harmonic oscillator potential:

$$\hat{H}_{oHF} = \frac{1}{2M} \cdot \sum_{n,j=1}^{A,3} \hat{p}_{nj}^2 + \frac{M\omega_1^2}{2} \cdot \sum_n x_n^2 + \frac{M\omega_2^2}{2} \cdot \sum_n y_n^2 + \frac{M\omega_3^2}{2} \cdot \sum_n z_n^2, \qquad (29)$$

The solution of Eq. (22) for $\hat{h}_o$ in Eq. (29) is determined [13,61-66] using a canonical or unitary transformation to eliminate the cross terms $y_n p_{nz}$ and $z_n p_{ny}$ in Eq. (22), and obtain the following transformed harmonic oscillator Hamiltonian:

$$\hat{H}_{crHF} = \frac{1}{2M} \cdot \sum_{n,j=1}^{A,3} p_{nj}^2 + \frac{M\omega_1^2}{2} \cdot \sum_n x_n^2 + \frac{M\alpha_2^2}{2} \cdot \sum_n y_n^2 + \frac{M\alpha_3^2}{2} \cdot \sum_n z_n^2 \qquad (30)$$

and the corresponding intrinsic (rotating-frame) or the cranked HF ground-state energy eigenvalue:

$$E_{crHF} = \hbar\omega_1 \Sigma_1 + \hbar\alpha_2 \Sigma_2 + \hbar\alpha_3 \Sigma_3 \qquad (31)$$

where:

$$\alpha_2^2 \equiv \omega_+^2 + \omega^2 + \sqrt{\omega_-^4 + 4\omega^2 \omega_+^2}, \qquad \alpha_3^2 \equiv \omega_+^2 + \omega^2 - \sqrt{\omega_-^4 + 4\omega^2 \omega_+^2} \qquad (32)$$

$$\omega_+^2 \equiv \frac{\omega_2^2 + \omega_3^2}{2}, \qquad \omega_-^2 \equiv \frac{\omega_2^2 - \omega_3^2}{2}, \qquad \Sigma_k \equiv \sum_{n_k=0}^{n_{kf}} (n_k + 1/2) \qquad (33)$$

where $n_{kf}$ is the number of oscillator quanta in the $k^{th}$ direction at the Fermi surface. Using this solution, we then obtain:

$$\langle \Phi_{crHF} | \hat{J} | \Phi_{crHF} \rangle = \omega \cdot \left[ M \mathcal{S}_+^o + \frac{4\hbar}{\alpha_2^2 - \alpha_3^2} (\alpha_3 \Sigma_3 - \alpha_2 \Sigma_2) \right] \qquad (34)$$

To satisfy the rigid-flow constraint in Eq. (23), we require in Eq. (34) the condition:

$$\langle \Phi | \sum_n \hat{p}_{n2}^2 | \Phi \rangle = \langle \Phi | \sum_n \hat{p}_{n3}^2 | \Phi \rangle \qquad \Rightarrow \qquad \alpha_2 \Sigma_2 = \alpha_3 \Sigma_3 \qquad (35)$$

so that the second term on the right-hand-side of Eq. (34) vanishes, and Eq. (34) reduces to:

$$\langle \Phi_{crHF} | \hat{J} | \Phi_{crHF} \rangle = \omega \cdot M \mathcal{S}_+^o \qquad (36)$$

Eq. (36) is valid for any value of the angular velocity $\omega$, which is determined as a function of the angular-momentum quantum number $J$ using the usual cranking model condition in Eq. (17)

to satisfy Eq. (23). Eq. (35) is the isotropic velocity distribution condition used by Ripka-Blaizot-Kassiss [13] as a self-consistency condition on the oscillator-potential frequencies, and yielded the rigid-flow value for the moment of inertia in their study using the conventional cranking model.

Having chosen the ratio of the oscillator frequencies as in Eq. (35), we then determine (as in [13]) all three oscillator frequencies by requiring that $\hat{h}_o$ in Eqs. (26) and (30) to approximate a HF mean-field Hamiltonian. Therefore, we choose the frequencies to minimize the energy $E_{crHF}$ in Eq. (31) subject the constant nuclear-quadrupole-volume condition:

$$\langle x^2 \rangle \cdot \langle y^2 \rangle \cdot \langle z^2 \rangle = c_o \tag{37}$$

where $\langle x_k^2 \rangle \equiv \langle \Phi_{crHF} | \sum_n x_{nk}^2 | \Phi_{crHF} \rangle$ ($k = 1,2,3$) and $c_o$ is a constant. This minimization yields a self-consistency between the shapes of nuclear equi-potential and equi-density surfaces [12,61,62,63].

The conditions in Eqs. (35) and (37) render the frequencies $\omega_1$, $\alpha_2$, $\alpha_3$, and hence the intrinsic energy $E_{crHF}$ in Eq. (31) constants independent of $\omega$. Therefore, since $E_{crHF}$ is minimized at $\omega = 0$, it remains so at all values of $\omega$. Nevertheless, $\omega_2$ and $\omega_3$ decreases and increases respectively with $\omega$ according to:

$$2\omega_+^2 = \alpha_2^2 + \alpha_3^2 - 2\omega^2, \qquad 4\omega_-^4 = \left(\alpha_2^2 - \alpha_3^2\right)^2 - 16\omega^2 \omega_+^2 \tag{38}$$

and equalize at $\omega = (\alpha_2 \pm \alpha_3)/2$ when the nucleus becomes axially symmetric about the rotation axis, at the cut-off maximum angular momentum $J = \Sigma_3 - \Sigma_2$. Therefore, starting from a prolate, oblate, or triaxial shape in its ground state, the shape of the nucleus changes as $\omega$ increases eventually becoming axially symmetric at the maximum angular momentum. The angular velocity $\omega$ is given by the solution of Eq. (36), which is evaluated to be:

$$J = \omega \cdot \left[ \frac{\Sigma_2}{\alpha_2} + \frac{\Sigma_3}{\alpha_3} + \frac{4\omega^2}{\alpha_2^2 - \alpha_3^2} \left( \frac{\Sigma_2}{\alpha_2} - \frac{\Sigma_3}{\alpha_3} \right) \right] \tag{39}$$

The excited-state and excitation energies of the HF ground-state rotational-band are given by:

$$E_{JcrHF} = \langle \Phi_{crHF} | \hat{H}_{oHF} | \Phi_{crHF} \rangle = \langle \Phi_{crHF} | \hat{H}_{crHF} | \Phi_{crHF} \rangle + \hbar \omega \cdot J \tag{40}$$

$$\Delta E_{JHF} = E_{JcrHF} - E_{J=0\,crHF} = E_{J=0crHF} + \hbar \omega \cdot J - E_{J=0\,crHF} = \hbar \omega \cdot J \tag{41}$$

where we have used the fact that the cranked self-consistent (HF) energy in Eq. (31) is a constant independent of $J$. It follows from Eqs. (36) and (41) that:

$$\Delta E_{JHF} = \hbar \omega \cdot J = \frac{J^2}{M \mathcal{S}_+^o} \equiv \frac{J^2}{2\mathcal{S}}$$

where $\mathcal{S} \equiv \dfrac{M \mathcal{S}_+^o}{2}$. This result shows that the independent-particle self-consistent cranking model predicts a moment of inertia that is on-half of the rigid-flow moment.

Fig 1 shows that the excitation energy $\Delta E_J$ predicted for $^{20}_{10}Ne$ by Eq. (41) is nearly the same as that predicted by the conventional cranking model (*CCRM*). The slight difference between them arises from the different moments of inertia in the two models (refer to Section 3 for a discussion of this difference). These predicted excitation energies deviate significantly from the measured excitation energy for $J > 4$, the experimentally observed moment of begins to increase from its nearly one-half of the rigid-flow value at about $J = 4$.

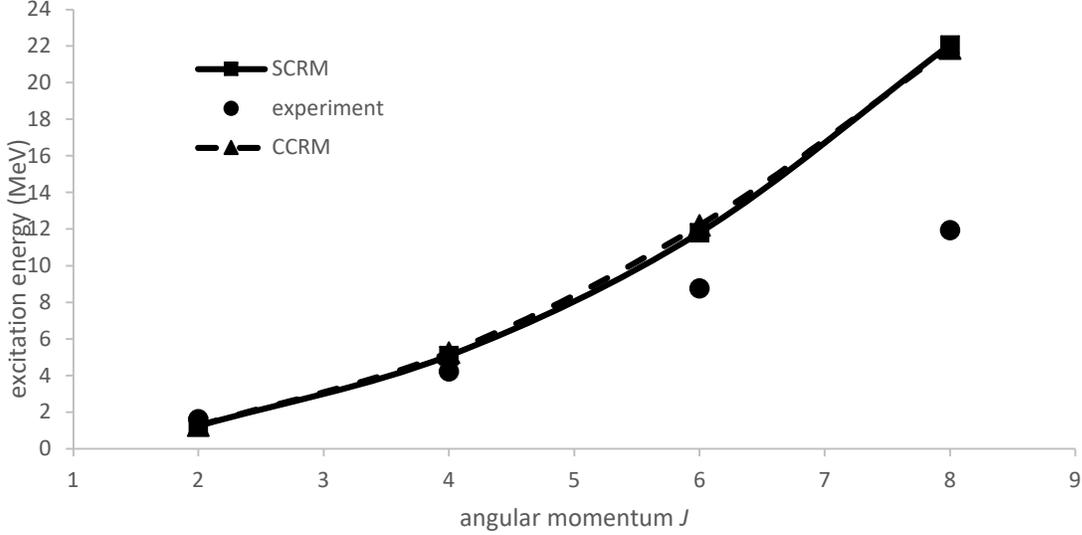

Fig 1: excitation energy versus J in Ne-20

## 5. Solving Eq. (25) for $^{20}_{10}Ne$ using Tamm-Dancoff approximation

To remove the discrepancy, shown in Fig 1, between the excitation energy predicted by the microscopic cranking HF (mean field) model and the measured excitation energy, we include in the microscopic ideal rotor model the residual two-body interaction and square of the angular momentum by solving in this section the model Eq. (25), with $\omega$ given by Eq. (36) or (39), using the Tamm-Dancoff method. Since the Hamiltonian $\hat{H}$ in Eq. (25) is purely intrinsic, there is no Goldstone rotational mode in an RPA application to Eq. (25)[9].

In the Tamm-Dancoff method [7,11,17], the wavefunction $|\Phi\rangle$ in Eq. (25) is expanded in the cranked HF particle-hole states as follows:

$$|\Phi\rangle = \sum_{mj} C_{mj} a_m^\dagger a_j |crHF\rangle \equiv \sum_{mj} C_{mj} |m\, j^{-1}\rangle \qquad (42)$$

where $|crHF\rangle$ is the cranked HF ground state and $|m\, j^{-1}\rangle$ is a cranked HF particle-hole state, where *m* refers to a particle state $|\varphi_m\rangle$ (given in Eq. (27)) above the Fermi surface, and *j* refers to a hole state $|\varphi_j\rangle$ at or below the Fermi surface. Substituting $|\Phi\rangle$ in Eq. (42) into Eq. (25), using Eqs. (24) and (28), and multiplying the resulting equation on the left by $\langle nk^{-1}|$, we obtain:

$$(\bar{\varepsilon}_{nk} - E_{ex}) \cdot C_{nk} = q_o \cdot q_{nk} + j_o \cdot j_{nk} + \frac{1}{2M\mathcal{S}_+^o} \cdot \left( \sum_m B_{nm} \cdot C_{mk} - \sum_j B_{kj} \cdot C_{nj} \right) \quad (43)$$

where:

$$\bar{\varepsilon}_{nk} \equiv \varepsilon_{nk} + \frac{1}{2M\mathcal{S}_+^o} \cdot \sum_{l=1}^{A} \left[ -(\hat{j}^2)_{ll} + \sum_{k=1}^{A} \hat{j}_{lk} \cdot \hat{j}_{kl} \right] + \frac{1}{2M\mathcal{S}_+^o} \cdot <\hat{J}>^2 \quad (44)$$

$$\varepsilon_{nk} \equiv \varepsilon_n - \varepsilon_k, \qquad E_{ex} \equiv E_{gs} - E_{crHF} \quad (45)$$

$$q_o \equiv \chi \cdot \sum_{mj} C_{mj} q_{mj}^*, \qquad j_o \equiv \frac{1}{2M\mathcal{S}_+^o} \cdot \sum_{mj} C_{mj} \cdot j_{mj}^* \quad (46)$$

$$B_{mn} \equiv (\hat{j}^2)_{mn} - 2\sum_{k=1}^{A} \hat{j}_{mk} \cdot \hat{j}_{kn}, \qquad B_{kj} \equiv (\hat{j}^2)_{kj} - 2\sum_{l=1}^{A} \hat{j}_{kl} \cdot \hat{j}_{lj} \quad (47)$$

---

[9] The results from an RPA solution of Eq. (25) seem to be similar to those from the Tamm-Dancoff solution. These RPA results are not reported here because they are preliminary and the solutions of the RPA equations require making additional approximations and are more involved.

$\varepsilon_{nk}$ in Eq. (45) is the particle-hole excitation energy. $Q_o$ and $j_o$ in Eq. (46) are constants, which can be readily determined.

The last term on the right-hand-side of Eq. (43) has a small effect on the results for $^{20}_{10}Ne$ and is therefore neglected in this article. Eq. (43) then gives:

$$C_{nk} = \frac{q_o \cdot q_{nk} + j_o \cdot j_{nk}}{\bar{\varepsilon}_{nk} - E_{ex}} \quad (48)$$

Substituting Eq. (48) into Eq. (46) and solving the resulting equation, we obtain the following approximate dispersion relation for the excitation energy $E_{ex}$ in Eq. (45):

$$1 - \chi \cdot \sum_{nk} \frac{q_{nk}^* \cdot q_{nk}}{\bar{\varepsilon}_{nk} - E_{ex}} - \frac{1}{M\mathcal{S}_+^o} \cdot \sum_{nk} \frac{j_{nk}^* \cdot j_{nk}}{\bar{\varepsilon}_{nk} - E_{ex}}$$
$$+ \frac{\chi}{M\mathcal{S}_+^o} \cdot \sum_{nk} \frac{q_{nk} \cdot q_{nk}^*}{\bar{\varepsilon}_{nk} - E_{ex}} \cdot \sum_{mj} \frac{j_{mj}^* \cdot j_{mj}}{\bar{\varepsilon}_{mj} - E_{ex}} - \frac{\chi}{M\mathcal{S}_+^o} \cdot \sum_{nk} \frac{q_{nk} \cdot j_{nk}^*}{\bar{\varepsilon}_{nk} - E_{ex}} \cdot \sum_{mj} \frac{q_{mj}^* \cdot j_{mj}}{\bar{\varepsilon}_{mj} - E_{ex}} = 0 \quad (49)$$

Eq. (49) has the usual property that each, except the smallest, of the $E_{ex}$ roots of Eq. (49) is located between a pair of the particle-hole excitation energies $\bar{\varepsilon}_{nk}$. The smallest $E_{ex}$ root of Eq. (49) is located below the smallest $\bar{\varepsilon}_{nk}$.

The rotational-band excited-state energy is given by:

$$E_J = \langle \Phi | H_o | \Phi \rangle = \langle \Phi | H | \Phi \rangle + \langle \Phi | \frac{\hat{J}^2}{2M\mathcal{S}_+^o} | \Phi \rangle$$
$$= E_{gs} + \hbar\omega \cdot J + \langle \Phi | \frac{(\hat{J}^2)_{res}}{2M\mathcal{S}_+^o} | \Phi \rangle = E_{crHF} + E_{ex} + \hbar\omega \cdot J + \langle \Phi | \frac{(\hat{J}^2)_{res}}{2M\mathcal{S}_+^o} | \Phi \rangle \quad (50)$$

where we have used Eqs. (20), (27) and (45). The rotational-band excitation energy is given by:

$$\Delta E_J = E_J - E_{J=0} = E_{ex} + \hbar\omega \cdot J + \langle\Phi|\frac{(\hat{J}^2)_{res}}{2M\mathcal{S}_+^o}|\Phi\rangle \tag{51}$$

since $E_{crHF}(J) = E_{crHF}(J=0) + \hbar\omega \cdot J$ (refer to Eq. (41)). Using Eqs. (24) and (42), we obtain:

$$\langle\Phi|(\hat{J}^2)_{res}|\Phi\rangle = \sum_{nmj} C_{nk}^* B_{nm} \cdot C_{mk} - \sum_{nkj} C_{nk}^* B_{kj} \cdot C_{nj}$$
$$+ \sum_{l=1}^{A}\left[(\hat{j}^2)_{ll} - \sum_{k=1}^{A} \hat{j}_{lk} \cdot \hat{j}_{kl}\right] - <\hat{J}>^2 + 2 \cdot \sum_{nk} C_{nk}^* \cdot \hat{j}_{nk}^* \cdot \sum_{mj} C_{mj} \cdot j_{mj} \tag{52}$$

## 6. Predictions of microscopic ideal rotor model for $^{20}_{10}Ne$

We have solved the microscopic ideal rotor-model (*MROTM*) dispersion relation in Eq. (49) in a closed form for the $^{20}_{10}Ne$ nucleus as follows. For the cranked self-consistent mean-field harmonic oscillator model given by Eq. (30), the mean-field ground state for $^{20}_{10}Ne$ has the prolate nucleon configuration: $(000)^4(100)^4(010)^4(001)^4(002)^4$, and the orbitals $(200),(020)$, $(110),(101),(011)$ and higher-lying orbitals are unoccupied. In Eq. (49), higher particle-hole excitations have higher particle-hole excitation energies and smaller associated matrix elements, and hence have small contributions in Eq. (49). Therefore, for simplicity, we restrict, in Eq. (44), the particle-hole excitations to the partially-filled valence shell with the total oscillator quantum number 2. The orbitals $(200),(020),(110),(101),(011)$ group into two sets of degenerate levels with respect to the orbital $(002)$. The first set has two degenerate levels each with the excitation energy: $\varepsilon_1 \equiv \varepsilon_{011} - \varepsilon_{002} = \varepsilon_{101} - \varepsilon_{002} = 0.4228\hbar\omega_o$. The second set has three degenerate levels each with the excitation energy: $\varepsilon_2 \equiv \varepsilon_{020} - \varepsilon_{002} = \varepsilon_{200} - \varepsilon_{002} = \varepsilon_{110} - \varepsilon_{002} = 0.8456\hbar\omega_o$. Therefore, each summation in Eq. (49) has only two terms, one term containing $\varepsilon_1$ and the second term containing $\varepsilon_2$. In this equation, we can ignore, for simplicity, $E_{ex}$ compared to $\bar{\varepsilon}_2$ since $\bar{\varepsilon}_2$ is much larger than $E_{ex}$, and the resulting equation is solved analytically to obtain the solution:

$$E_{ex} = \bar{\varepsilon}_1 - \frac{1}{2a} \cdot \left(b \pm \sqrt{b^2 - 4a\cdot c}\right) \tag{53}$$

where:

$$a \equiv \bar{\varepsilon}_2^2 - \bar{\varepsilon}_2 \cdot \left(\chi \cdot q_2^2 + \frac{j_2^2}{M\mathcal{S}_+^o}\right) + \frac{\chi}{M\mathcal{S}_+^o} \cdot \left(q_2^2 \cdot j_2^2 - |\Lambda_2|^2\right) \tag{54}$$

$$b \equiv \bar{\varepsilon}_2^2 \cdot \left(\chi \cdot q_1^2 + \frac{j_1^2}{M\mathcal{S}_+^o}\right) + \frac{\chi \cdot \bar{\varepsilon}_2}{M\mathcal{S}_+^o} \cdot \left(q_1^2 \cdot j_2^2 + q_2^2 \cdot j_1^2 - \Lambda_1\Lambda_2^* - \Lambda_1^*\Lambda_2\right) \tag{55}$$

$$c \equiv \frac{\chi \cdot \bar{\varepsilon}_2^2}{M\mathcal{S}_+^o} \cdot \left(q_1^2 \cdot j_1^2 - |\Lambda_1|^2\right) \tag{56}$$

$$q_1^2 \equiv \sum_{mj}^{set1} q_{mj} q_{mj}^*, \quad q_2^2 \equiv \sum_{mj}^{set2} q_{mj} q_{mj}^*, \quad j_1^2 \equiv \sum_{mj}^{set1} j_{mj} j_{mj}^*, \quad j_2^2 \equiv \sum_{mj}^{set2} j_{mj} j_{mj}^* \tag{57}$$

$$\Lambda_1 \equiv \sum_{mj}^{set1} q_{mj}^* j_{mj}, \quad \Lambda_2 \equiv \sum_{mj}^{set2} q_{mj} j_{mj}^* \tag{58}$$

For the uni-axial rotation rotor-model case that we are considering in this article, the quantities $j_2$ and $q_2$ (for the second set of the levels) vanish. For this case, the dashed line with triangle symbols in Fig 2 shows $\Delta E_J$ predicted by Eqs. (51), (52) and (53) using the values $\varepsilon_1 = 0.1\hbar\omega_o$ and $\chi = 1.1$ that give the best fit to the measured excitation energies. This chosen value of $\varepsilon_1$ is smaller than the predicted HF value of $0.4228\hbar\omega_o$, and the chosen value of $\chi$ is much higher than those used in the literature for heavy nuclei. The overall agreement between the predicted and measured excitation energies is reasonably good but the convex shape of the yrast line above $J = 4$ is not predicted.

The convex shape was pointed out by Bohr-Mottelson [12], who attributed the shape to some unknown oscillatory physical phenomenon. There have been many predictions of the excitation-energy spectrum of $^{20}_{10}Ne$ using HF, SU(3), Sp(3,R), and phenomenological approaches [21,59,60,67-76]. Nearly all of these analyses predict an excitation-energy spectrum where the energy-level spacing increases monotonically with $J$ and a number of them predict a compressed spectrum. In [21,77], which used self-consistent deformed oscillator with $\vec{l}\cdot\vec{s}$ coupling and without any two-body interaction, the predicted excitation energy follows a straight line up to $J = 6$ and is lower than the measured excitation energy by as much as 2 $MeV$ at $J = 6$ and $J = 8$. The smaller predicted energy spacing between $J = 6$ and $J = 8$ relative to that between $J = 4$ and $J = 6$ is achieved by assuming that the oblate aligned state at $J = 8$ is rotating about the rotation axis, which at $J = 8$ is also the symmetry axis. The model does not predict decreasing excitation-energy spacing with $J$ at $J = 4$ and 6.

We surmise that the decrease in the excitation-energy spacing with $J$ is caused by the effects, among others, of a 3-D rotation and fluctuations in the angular momentum components and their interaction with fluctuations and vibrations in the mean-field potential generated by the residual quadrupole-quadrupole interaction. To determine qualitatively if this is the case, we have simulated 3-D rotation effects in the microscopic ideal rotor model as follows. These effects are from rotations about the $y$ and $z$ axes in addition to that about the $x$-axis. As mentioned above, the quantities $j_2$ and $q_2$ (for the second set of the cranked degenerate levels) in Eq. (57) vanish because we are considering a rotation along only the $x$-axis. For a 3-D rotation, these quantities would not vanish. To simulate this 3-D rotation effect, we have set in Eq. (57) $j_2 = \alpha_j \cdot j_1$, $q_2 = \alpha_q \cdot q_1$ and we have chosen $\alpha_j = -0.9$, and $\alpha_q = 1.0$. The solid line with square symbols in Fig 2 shows the excitation energy predicted by this model for $\varepsilon_1 = 0.1\hbar\omega_o$, $\varepsilon_2 = 1.0\hbar\omega_o$, $\chi = 0.78$ [10]. The predicted excitation energy agrees fairly well with the measured data (and differs markedly in shape from the excitation energy shown by the dashed line predicted by the one-set of degenerate-level case, i.e., for zero values of $j_2$ and $q_2$).

It is noted that the decreasing excitation-energy spacing with $J$ (i.e., the lowering of the yrast states) at $J = 4, 6,$ and $8$ in our model is caused by transverse fluctuations in the stable rotation along the $x$-axis of the stable HF mean field in Eqs. (26), (40), and (41), and by variations or vibrations in the HF mean field. These fluctuations and vibrations are described by the residual $(\hat{J}^2)_{res}$ part of the square of the angular momentum and by the residual two-body interaction $Q^\dagger \cdot Q$ in Eq. (25), and by the resulting variations in particle-hole matrix elements, where $j_1^2$ in Eq. (57) decreases to zero at $J = 8$ from its maximum value at $J = 0$, and $q_1^2$ in Eq. (57) increases from zero at $J = 0$ to its maximum value at $J = 8$[11].

The somewhat large value (-0.9) of $\alpha_j$ used in the Tamm-Dancoff analysis to obtain a fit to the measured excitation energy would imply a somewhat large transverse (i.e., $y$ and $z$) components of the angular momentum. This would in turn imply an effective rotation along an axis other than either of the principal axes of the mean-field potential, i.e., a tilted-axis rotation as studied in [26] using conventional cranking model, in contrast to precessing or wobbling of a high mean-field angular momentum along the $x$-axis studied in [12,27,28,80,81] using conventional cranking model. Note that for $^{20}_{10}Ne$, the angular momentum values 4, 6, and 8 are considered high since the ground-state rotational band terminates at $J = 8$ at which all the valence nucleons have aligned their angular momenta along the rotation $x$-axis. The seemingly tilted-axis rotation predicted by the simulated 3-D rotation in the model appears to be at odds with the conclusion arrived at in [26] that, in even-even nuclei, tilted-axis rotation cannot occur if the nucleus is axially symmetric in its ground state. However, the above predictions of the ideal rotor model, including the apparent tilted-axis rotation, will be re-examined when the model is generalized to include genuine a 3-D quantum rotation.

Fig 3 shows that the quadrupole moment predicted by the microscopic ideal rotor model and conventional cranking model for $^{20}_{10}Ne$ agrees reasonably well with the measured quadrupole moment in view of measurement uncertainties.

The results of an application of the RPA method for the one-set of degenerate-level case is found to be similar to those from the Tamm-Dancoff method. For the two-set of degenerate-level case with the simulated 3-D rotation, the RPA dispersion equation is complex and we have not yet found a converged first-order RPA solution to it. However, for this case, we expect RPA method to also yield results similar to those of the Tamm-Dancoff method.

---

[10] Such shifts in the single-particle energies relative to the cranked self-consistent oscillator values may be generated by the addition of $\vec{l}\cdot\vec{s}$ coupling interaction to the mean-field potential in Eq. (29) [21,77]. $l^2$ and pairing interactions are known [13,21,78,79] not to be important in $^{20}_{10}Ne$. Also the choice of more realistic values of the model parameters $\varepsilon_1$, $\varepsilon_2$, and $\chi$ may become possible when $\vec{l}\cdot\vec{s}$ coupling interaction and nucleon valence states in the higher unoccupied shells are included in the model.

[11] This perturbation of the stable rotation along the $x$-axis described by the cranked HF mean field is similar to the Goldstone rotational mode or wobbling excitation studied in the framework of the cranking model plus RPA, which describes a mode with zero excitation energy in the principal-axis and tilted-axis rotations [26,27,28,80,81].

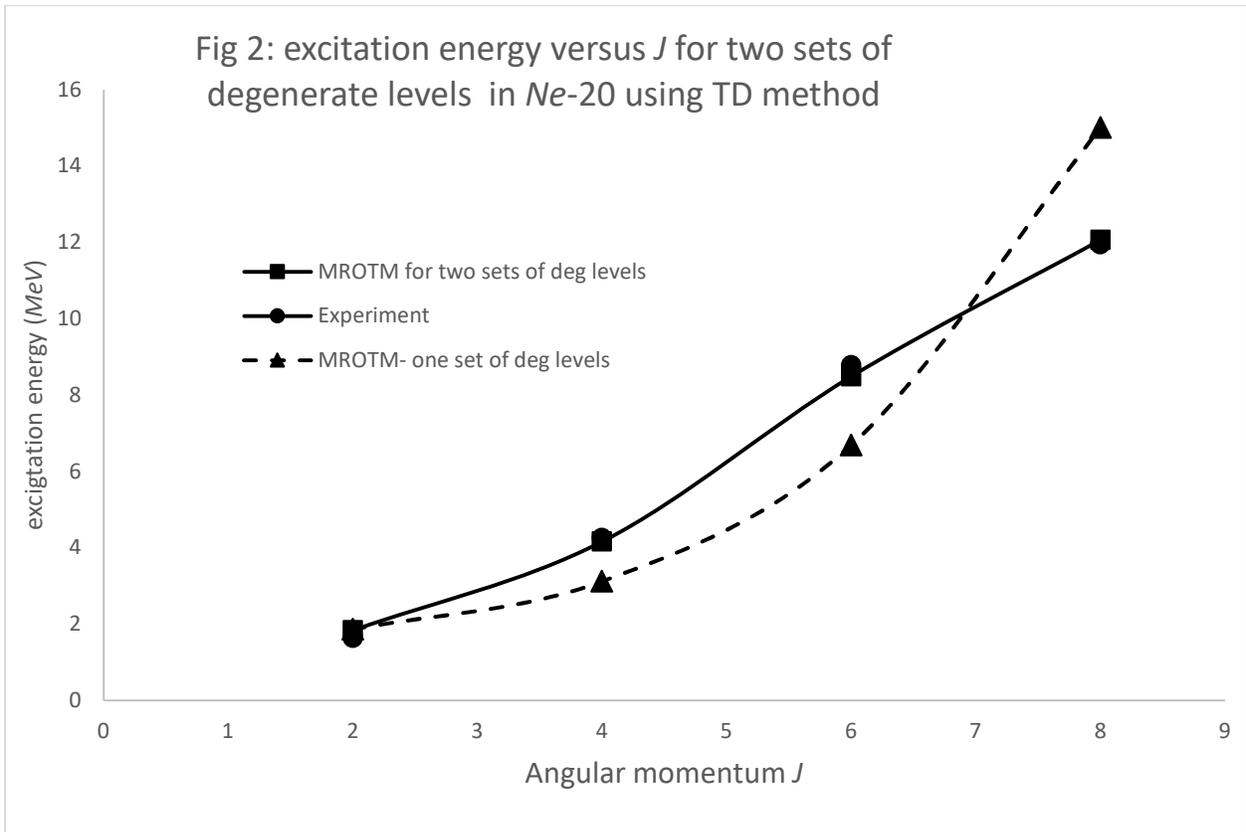

Fig 2: excitation energy versus *J* for two sets of degenerate levels in *Ne*-20 using TD method

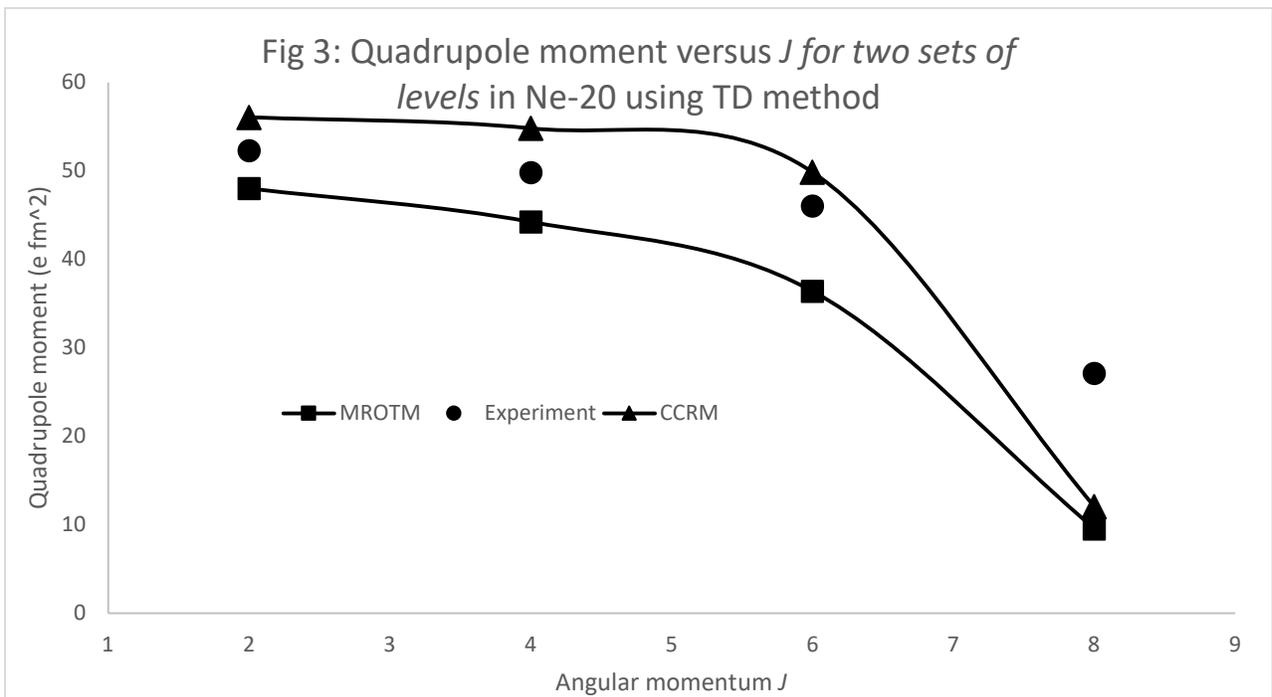

Fig 3: Quadrupole moment versus *J for two sets of levels* in Ne-20 using TD method

## 7. Concluding remarks

In this article, a microscopic quantum ideal rotor model is derived by applying a rotation operator to the nuclear ground-state (or a band-head) wavefunction describing a deformed nucleon distribution. The rotation angle in the rotation operator is chosen to specify the orientation of the quadrupole tensor component of the deformed nucleon distribution, and hence is considered to be a function of the nucleon co-ordinates. The rotation operator transforms the nuclear Hamiltonian into a rotor-model intrinsic Hamiltonian when the rotated wavefunction is substituted into the nuclear Schrodinger equation. The intrinsic Hamiltonian becomes purely (ideally) intrinsic, i.e., it becomes independent of angular momentum operator, when the rotation angle is chosen to describe a rigid-flow nucleon velocity field and canonically conjugate to the angular momentum operator. The intrinsic Hamiltonian resembles the intrinsic Hamiltonian in the Bohr's rotational model but with the rigid-flow kinematic moment of inertia instead of an arbitrary dynamic moment of inertia. It is argued that the intrinsic Hamiltonian is distinct from that in the Bohr's rotational model, and that the ideal rotor model is more appropriate for the description of the rotational motion in at least the light nuclei than the Bohr's rotational model. (For the case of the center-of-mass motion, the rotor model predicts the correct mass.)

The ideal rotor-model Hamiltonian is then reduced to a self-consistent cranking-model Hamiltonian plus correction terms related to the residuals of the square of the angular momentum and a two-body interaction using Hartree-Fock and second-quantization methods. It is shown that the conventional cranking model Hamiltonian is not purely intrinsic because it does not use a rigid-flow kinematic moment of inertia. The derived self-consistent cranking model equation is solved analytically for a self-consistent deformed harmonic oscillator potential. The ideal rotor-model Schrodinger equation with the residual of the square of the angular momentum operator and a separable quadrupole-quadrupole two body residual interaction is then solved in the Tamm-Dancoff (particle-hole) approximation using the cranked HF states as particle-hole basis states.

For $^{20}_{10}Ne$, the ideal rotor model predicts reasonably well the overall excitation energy when the interaction strength and the HF cranked particle-hole excitation energies are adjusted, but the convex shape of the yrast line (i.e., the decrease in the excitation-energy spacing with the angular momentum) is not predicted. However, when the effects of a 3-D rotation are simulated in the model, a close fit to the empirical excitation energy including its convex shape is obtained. It is thereby concluded that the convex yrast shape is the result of a tilted-axis rotation, i.e., a rotation along an axis other than a principal axis of the mean-field potential. The quadrupole moment is also reasonably well predicted.

In a future article, we will generalize the microscopic quantum ideal rotor model for uni-axial rotation to 3-D rotation and analyze the impact of this rotation on the rotational-band excitation energy.